\documentclass[conference,final]{IEEEtran}
\IEEEoverridecommandlockouts
\usepackage{cite}
\usepackage{amsmath,amssymb,amsfonts}
\usepackage{algorithmic}
\usepackage{graphicx}
\usepackage{textcomp}
\usepackage{xcolor}
\def\BibTeX{{\rm B\kern-.05em{\sc i\kern-.025em b}\kern-.08em
    T\kern-.1667em\lower.7ex\hbox{E}\kern-.125emX}}

\usepackage{glossaries}
\usepackage{hyperref}
\usepackage{cleveref}
\usepackage{xspace}
\usepackage{soul}
\usepackage{siunitx}
\usepackage{multirow}
\usepackage{url}

\newacronym{ML}{ML}{machine learning}
\newacronym{TCB}{TCB}{trusted computing base}
\newacronym{DIFT}{DIFT}{dynamic information flow tracking}
\newacronym{SoC}{SoC}{system-on-chip}
\newacronym{OS}{OS}{operating system}
\newacronym{PE}{PE}{processing element}
\newacronym{FHE}{FHE}{fully homomorphic encryption}
\newacronym{CPU}{CPU}{central processing unit}
\newacronym{GPU}{GPU}{graphics processing unit}
\newacronym{TEE}{TEE}{trusted execution environment}
\newacronym{REE}{REE}{rich execution environment}
\newacronym{DMA}{DMA}{direct memory access}
\newacronym{MLaaS}{MLaaS}{machine learning as a service}
\newacronym{HDL}{HDL}{hardware description language}
\newacronym{MAC}{MAC}{multiply-and-accumulate}
\newacronym{WNS}{WNS}{worst negative slack}
\newacronym{ISA}{ISA}{instruction set architecture}

\newcommand{\circnum}[1]{\raisebox{.5pt}{\textcircled{\raisebox{-.9pt} {#1}}}\xspace}

\newcommand{\design}{\textsf{Dolma}}

\begin{document}

\title{Data-Oblivious ML Accelerators using Hardware Security Extensions}

\author{\IEEEauthorblockN{Hossam ElAtali}
\IEEEauthorblockA{\textit{University of Waterloo} \\
hossam.elatali@uwaterloo.ca}
\and
\IEEEauthorblockN{John Z. Jekel}
\IEEEauthorblockA{\textit{University of Waterloo} \\
jzjekel@uwaterloo.ca}
\and
\IEEEauthorblockN{Lachlan J. Gunn}
\IEEEauthorblockA{\textit{Aalto University} \\
lachlan@gunn.ee}
\and
\IEEEauthorblockN{N. Asokan}
\IEEEauthorblockA{\textit{University of Waterloo} \\
asokan@acm.org}
}

\maketitle
\thispagestyle{plain}
\pagestyle{plain}

\begin{abstract}
Outsourced computation can put client data confidentiality at risk. Existing solutions are either inefficient or insufficiently secure: cryptographic techniques like fully-homomorphic encryption incur significant overheads, even with hardware assistance, while the complexity of hardware-assisted trusted execution environments has been exploited to leak secret data. 

Recent proposals such as BliMe and OISA show how \gls{DIFT} enforced in hardware can protect client data efficiently. They are designed to protect CPU-only workloads. However, many outsourced computing applications, like machine learning, make extensive use of accelerators.

We address this gap with \design{}, which applies \gls{DIFT} 
to the Gemmini matrix multiplication accelerator, \emph{efficiently} guaranteeing client data confidentiality, \emph{even in the presence of malicious/vulnerable software and side channel attacks} on the server. We show that accelerators can allow \gls{DIFT} logic optimizations that significantly reduce area overhead compared with general-purpose processor architectures. \design{} is integrated with the BliMe framework to achieve end-to-end security guarantees.
We evaluate \design{} on an FPGA using a ResNet-50 DNN model and show that it incurs low overheads for large configurations ($4.4\%$, $16.7\%$, $16.5\%$ for performance, resource usage and power, respectively, with a 32x32 configuration).
\end{abstract}

\glsresetall

\section{Introduction}

In traditional outsourced computation settings, clients must trust the providers of the computation service with their data. For example, a client wishing to classify an image using a service provider's \gls{ML} model must send the image to the service provider for \gls{ML} inference. This poses a confidentiality risk to client data and can even prevent clients from using the service due to legal privacy requirements. One solution to this is cryptographic methods such as \gls{FHE}, which allow processing on encrypted data. This guarantees confidentiality because the data is never decrypted on the service provider's side. However, \gls{FHE} suffers from excessive overheads. Even when using dedicated hardware accelerators such as CraterLake~\cite{craterlake2022}, \gls{FHE} performance is orders of magnitude worse than native computation on a \gls{CPU}, and even more in comparison with dedicated \gls{ML} accelerators.
Another solution is secure hardware mechanisms such as \glspl{TEE} and confidential computing. Examples include Intel SGX~\cite{intelSGXExplained2016}, Arm TrustZone~\cite{demystifyingTrustzone} and NVIDIA Confidential Computing~\cite{ConfidentialComputeNVIDIA}. These rely on a root-of-trust in hardware to provide a trusted and isolated environment (a ``\gls{TEE}'') for execution of client code. Programs, including the \gls{OS}, executing outside the \gls{TEE} are said to be in the ``\gls{REE}'' and are not allowed to view or modify data within the \gls{TEE}.
However, all existing commercial \gls{TEE} solutions are vulnerable to side-channel attacks and consider them out-of-scope. Furthermore, in these solutions, programs executing inside the \glspl{TEE} are considered \emph{trusted} and can exfiltrate data outside the \gls{TEE}. As a result, \glspl{TEE} are not suitable when the client \emph{cannot trust the code} processing their data. This is common in two cases: 1) the client relies on an outsourced service where they do not have access to the code, and 2) the client knows the code but cannot guarantee that it is free of bugs, which can lead to run-time exploits, and side-channel vulnerabilities.

Prior work has attempted to solve these problems. OISA~\cite{oisa} provides \gls{CPU} extensions that can be used to taint secret data and ensure that any tainted data is not leaked through side channels. However, OISA requires the processing code to be trusted and is therefore only useful in preventing accidental side channel leakage. BliMe~\cite{blime24} takes this a step further by removing all trust in software, therefore protecting against intentional and accidental leakage either directly or via side channels. Unfortunately, both OISA and BliMe only apply to \gls{CPU} workloads and do not address accelerators. 

In this work, we extend secure data-oblivious outsourced computation to hardware accelerators. We are the first to propose a confidential computing platform for hardware accelerators that provides resistance against software bugs and side channels \emph{without} requiring any software to be part of the \gls{TCB}, including the software processing the secret data. We use hardware-based \gls{DIFT} to track secret client data (and any derivative of it) and enforce a security policy that aborts any attempt, intentional or otherwise, by software to leak this data. Our implementation is based on BliMe, which provides this feature for CPU-based workloads on the BOOM core~\cite{boom}. Furthermore, while implementing \gls{DIFT} in accelerators using mechanisms such as GLIFT~\cite{GLIFT} is possible, it introduces large overheads due to the unnecessarily small granularity (gate-level) at which taint is propagated. The predictable behavior of systolic arrays common in accelerator architectures presents unique optimization opportunities. We can safely propagate taint at a much higher granule leading to significantly reduced area overheads.%

Our contributions are the following:
\begin{itemize}
    \item We present \design{}, a minimal extension to matrix multiplication hardware accelerators that enables efficient \gls{DIFT} and guarantees confidentiality of secret client data, including against side-channel attacks and even in the presence of malicious software. We provide an implementation of \design{} on Gemmini~\cite{gemmini2021}, a flexible hardware accelerator integrated into the RISC-V Chipyard~\cite{chipyard} \gls{SoC} (\Cref{sec:impl}).
    \item Using realistic \gls{ML} workloads we show that our implementation has minimal overheads compared to the unsecured baseline, and orders-of-magnitude lower overheads compared to state-of-the-art cryptographic solutions with similar threat models ($4.4\%$ performance overhead with a 32x32 configuration; \Cref{sec:performance}).
    \item We evaluate the resource usage overhead and show that our implementation is scalable and feasible for accelerators with large dimensions (\Cref{sec:power}).
    \item We extend the existing BliMe F* model with a model of accelerator operations and provide a machine-checked proof of its security (\Cref{sec:security}).%
\end{itemize}

\section{Background}

\subsection{Side channels}\label{sec:sidechannels}
Side channels are unintended outputs of a system. They can be used to reveal information about the data being processed by the system, such as by measuring timing of different operations. For example, early RSA implementations had variable execution times, where the time to process each bit of the private key depended on whether the bit was a zero or a one~\cite{kocher1996timing}. This allowed an attacker to analyze the time taken to complete the operation and therefore determine the number of `one' bits in the key, which can be used to recover the key.

Another timing side channel results from memory accesses whose addresses depend on secret data. This is because memory accesses can modify the state of internal CPU caches, which can then be probed to leak the secret data. A prominent example of a cache timing attack is Prime+Probe~\cite{primeProbe}. The attacker first primes the cache by filling all entries with their data and waits for the victim to perform the secret-dependent memory access. The victim's memory access evicts one of the attacker's entries from the cache. The attacker can then probe the cache by attempting to access an entry from each cache line. The cache lines that take longer to access have been evicted, and this can be used to leak the victim's secret.

\subsection{BliMe}
BliMe~\cite{blime24} guarantees client data confidentiality by tracking secret data and any derivative values using a \gls{DIFT} policy. The \gls{DIFT} policy is enforced by modifying the \gls{CPU} such that all instructions either propagate the taint (e.g., using a tainted operand in an add instruction will result in a tainted sum), or fault due to violating BliMe's security policy (e.g., by using tainted data as an address for a memory access or as a branching condition, or attempting to write tainted data to untrusted I/O).

BliMe's protocol is as follows. The client first initiates a handshake with the BliMe-enabled server and agrees with it on a secret session key. The client then encrypts its data with this key and sends it to the server where the processing software invokes a BliMe instruction to atomically decrypt and ``blind'' (i.e., taint) the data. The software can then process the data, but all processing is subject to the \gls{DIFT} policy described above. Any attempt to leak blinded data, including through side channels, is aborted by BliMe. This allows \emph{untrusted} software to operate on secret client data while its confidentiality is guaranteed solely by the hardware. Once processing is complete, the software invokes a BliMe instruction to atomically encrypt and unblind the result, and then sends it back to the client.

BliMe is implemented on the speculative out-of-order BOOM core~\cite{boom} and Chipyard SoC~\cite{chipyard}. Both BOOM and Chipyard are written in Chisel~\cite{chisel}, a \gls{HDL} embedded in Scala. BliMe uses memory tagging to differentiate between blinded and non-blinded data, with tags attached to registers, cache entries and physical memory: each 64 bits of data are augmented with an 8-bit tag. The additional memory requests required to read/write the tags between the last-level cache and physical memory introduces an average performance overhead of 8\%.

\subsection{Gemmini}\label{sec:bg:gemm}
Gemmini is a RISC-V accelerator that allows flexible configuration of the accelerator parameters, e.g., systolic array dimensions, data flow (weight stationary or output stationary), and scratchpad sizes. It is connected to the host CPU using the RoCC interface~\cite{chipyard}. It consists of a systolic array (a ``mesh'') for matrix multiplication, internal scratchpads for data storage, and components for performing activation functions such as ReLU. The circuit is controlled by three controllers: two for \gls{DMA} and one for execution. These controllers can act independently resulting in a decoupled access-execute architecture. The systolic array has two levels: the mesh consists of a 2D array of tiles, and each tile is a combinational 2D array of \glspl{PE}. The main computation performed by the systolic array is the following \gls{MAC} operation (for a 2x2 array):

\begin{equation}
C = A * B + D
\end{equation}

\begin{equation}\label{eq:elements}
\begin{pmatrix}
    c_{11} & c_{12}\\
    c_{21} & c_{22}
\end{pmatrix}
=
\begin{pmatrix}
    a_{11} & a_{12}\\
    a_{21} & a_{22}
\end{pmatrix}
\begin{pmatrix}
    b_{11} & b_{12}\\
    b_{21} & b_{22}
\end{pmatrix}
+
\begin{pmatrix}
    d_{11} & d_{12}\\
    d_{21} & d_{22}
\end{pmatrix}
\end{equation}
\vspace{1em}

Input matrices are loaded from memory and stored in the scratchpads, whose width corresponds to the width of the systolic array. The inputs then flow into the array, where each \gls{PE} performs a \gls{MAC} on a single element, e.g., $a_{11}*b_{11}+d_{11}$. The inputs must be delayed appropriately using registers to ensure the correct elements are \gls{MAC}ed in every cycle. The outputs are also delayed to ensure the resulting matrix is written into the scratchpads \emph{one row at a time}.

Gemmini supports two data flow modes: weight stationary and output stationary. In the former, the weights, $B$, are preloaded into the array and $A$ and $D$ flow in. In the latter, $D$ is preloaded and $A$ and $B$ flow in. With weight-stationary data flow, inputs A and D can be expanded to allow back-to-back computations on \emph{independent rows}, as shown in \Cref{eq:WS}. This is particularly useful in neural networks where multiple inputs, e.g., images, are given to the same model. The model weights, $B$, remain in place and the input vectors ($A_1$, $A_2$, $A_3$, etc.) are fed in as independent rows with no stalls between them. %

\begin{equation}\label{eq:WS}
\footnotesize
\begin{pmatrix}
    c_{11} & c_{12}\\
    c_{21} & c_{22}\\
    c_{31} & c_{32}\\
    ...    & ...   \\
    c_{K1} & c_{K2}
\end{pmatrix}
=
\begin{pmatrix}
    a_{11} & a_{12}\\
    a_{21} & a_{22}\\
    a_{31} & a_{32}\\
    ...    & ...   \\
    a_{K1} & a_{K2}
\end{pmatrix}
\begin{pmatrix}
    b_{11} & b_{12}\\
    b_{21} & b_{22}
\end{pmatrix}
+
\begin{pmatrix}
    d_{11} & d_{12}\\
    d_{21} & d_{22}\\
    d_{31} & d_{32}\\
    ...    & ...   \\
    d_{K1} & d_{K2}
\end{pmatrix}
\end{equation}
\vspace{1em}

\section{Assumptions \& Threat Model}

Our \gls{TCB} includes the BliMe CPU hardware, which we assume correctly enforces the BliMe \gls{DIFT} policy, and the accelerator hardware as well as the interconnect between them, whose data we assume cannot be sniffed or modified. We also assume that BliMe can securely perform remote attestation and manage client session keys and key-to-tag mappings.

\emph{All} software, including the \gls{OS} and any applications processing client data, is untrusted. We assume the adversary can control all software running on the server. As in~\cite{blime24}, we consider attacks requiring intrusive physical access, such as bus snooping or differential power analysis, to be out of scope.

\section{Design \& Implementation}\label{sec:impl}

\begin{figure}
    \centering
    \includegraphics[trim={6.5cm 5cm 6cm 6cm},clip,width=0.97\linewidth]{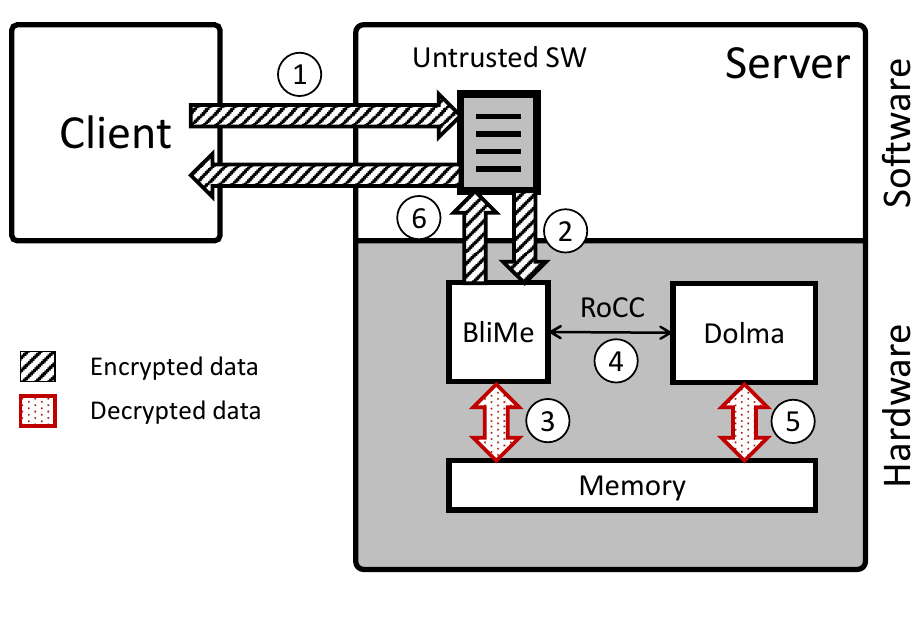}
    \caption{System Overview. The client encrypts and send their secret data to the untrusted software on the server~\circnum{1}, which calls BliMe's data import operation~\circnum{2}. BliMe decrypts-and-blinds the data, tagging it with the session-specific tag, and stores it in memory~\circnum{3}. The untrusted software can then use RoCC instructions~\circnum{4} to make \design{} operate on the blinded data. \design{} accesses the data using \gls{DMA} and enforces the \gls{DIFT} security policy~\circnum{5}. Once the processing is complete, the untrusted software can call BliMe's data export function to encrypt-and-unblind the data, and then send it back to the client~\circnum{6}.}
    \label{fig:overview}
\end{figure}

\subsection{Overview}
An overview of the system is shown in \Cref{fig:overview}. \design{} is connected to a BliMe CPU through the RoCC interface. The remote attestation performed by BliMe in the handshake step is augmented to include attestation of \design{} using an additional root of trust embedded in \design{}. This assures the client that the server contains a genuine BliMe \gls{CPU} connected to a genuine \design{} accelerator. The client encrypts its data using the session key shared with BliMe and sends it to the server. The processing software on the server can then call BliMe's data import instruction to atomically blind and decrypt the data. At this point, the secret client data is blinded in memory and can be read by \design{} for processing. Note that combining blinded and non-blinded data (e.g., an input image and a model, respectively) within \design{} is allowed and the result (e.g., the classification output) is blinded. Once processing is complete, the blinded result can be atomically unblinded and encrypted using BliMe's data export instruction and sent back to the client.

Our \gls{DIFT} policy is defined such that
if any of the following have the same non-zero tag, then the output row receives the same non-zero tag%
:
\begin{itemize}
  \item the corresponding input row $\mathbf{a}_i$,
  \item the corresponding input row $\mathbf{d}_i$,
  \item any weight matrix row $\mathbf{b}_j$.
\end{itemize}
If any two of these rows have differing non-zero tags, then the operation faults and does not emit any further output.

\gls{DIFT} is used by the main CPU to forbid software from using secret data in a way that affects memory access patterns, control flow, or I/O. The \gls{DIFT} policy above ensures that \design{} cannot be used to circumvent these restrictions. In the remainder of this section, we will present the challenges we encountered while implementing this policy in Gemmini and the optimizations we used to reduce overheads. Even though we discuss Gemmini in particular, these challenges and optimizations are applicable to a wide range of hardware accelerators as they concern generic commonly used mechanisms.

\subsection{Tag bits}
We extend the RoCC interface to include tag bits for the register values passed to Gemmini from the CPU. We also extend the memory interface and internal scratchpads to include the tag bits. The TLB is \emph{not} extended with tag bits as page table entries must not be blinded. We check all new entries into the TLB; if an entry is blinded, it is zeroed and a fault is raised.

We ensure the completeness of our modifications by using a \texttt{Blinded} data type~\cite{blime24}, a wrapper for untagged Chisel data bundles.  This wrapper causes the Chisel compiler to raise an error wherever non-BliMe-aware logic attempts to use tagged data. %
Upon detecting such a mismatch, we decide whether to 1) safely propagate the tags, or 2) check for a security violation and if one is detected, zero out the blinded value and raise a fault.

\subsection{RoCC commands}
Sending blinded instructions to the accelerator is prohibited because, otherwise, the type of operation performed by the accelerator, e.g., loading values or starting execution, would leak the values of the instructions. Furthermore, in Gemmini's case, RoCC commands pass two register values to Gemmini as `rs1' and `rs2', which are used either as pointers to data in memory or as opcodes for subfunctions. In both cases, blinded values for rs1 and rs2 are prohibited by the hardware because they can be leaked through the memory access patterns or by observing the subfunction performed by Gemmini, respectively. 

\subsection{DIFT in the systolic array}\label{sec:dift}
\begin{figure*}[ht]
    \centering
    \includegraphics[trim={0cm 2.5cm 0cm 2.5cm},clip,width=\linewidth]{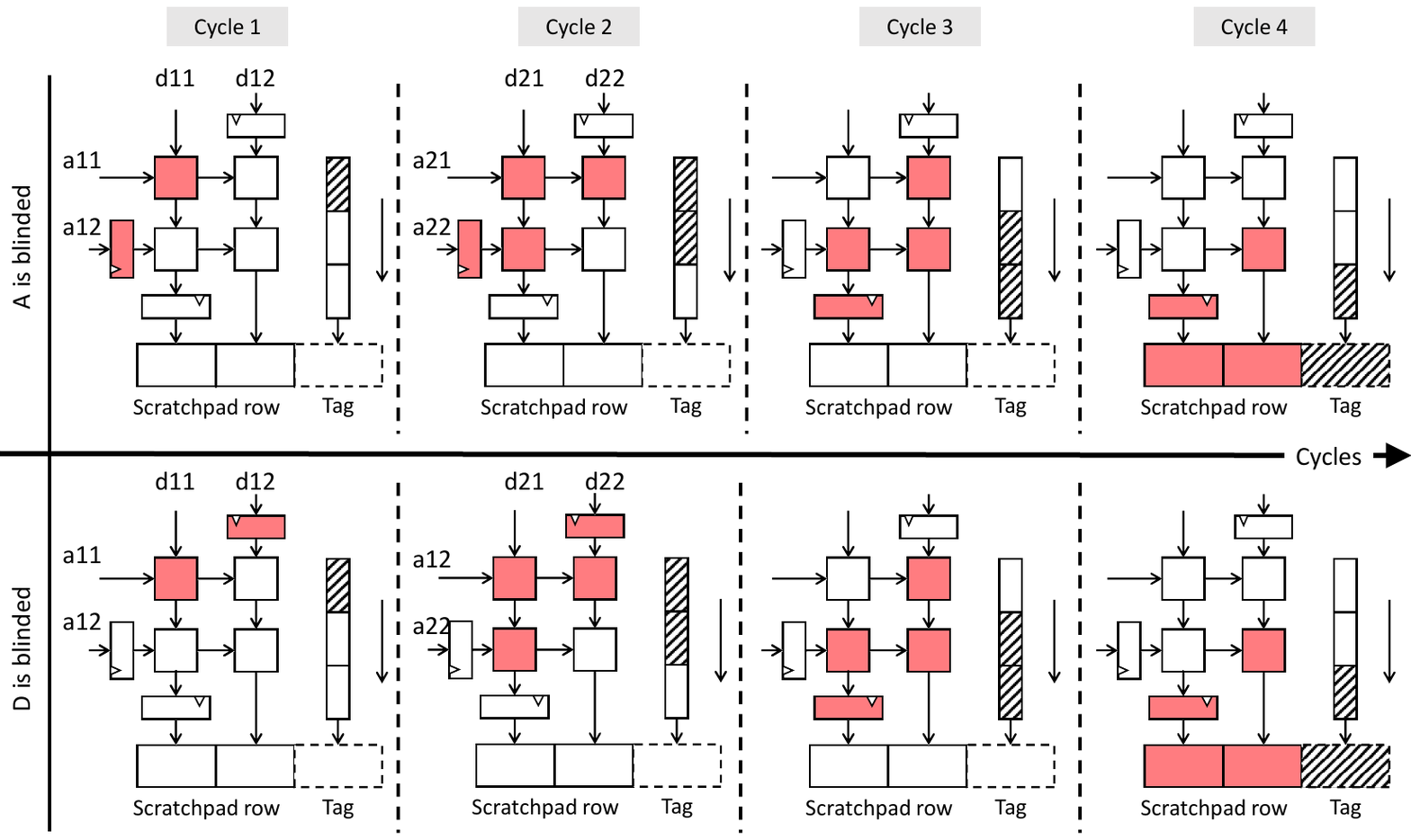}
    \caption{
    \gls{DIFT} in parallel for the weight-stationary data flow inside a 2x2 systolic array. The top half of the figure shows the case where $A$ is blinded. The bottom half shows the case where $D$ is blinded. The subfigures from left to right show how computation proceeds over successive cycle. Secret values are shown in red. Note that the tiles and registers themselves do not carry any additional logic. %
    The tags corresponding to the secret values are shown as striped. Output tags are calculated before input values enter the systolic array and propagate alongside the corresponding secret input and intermediate values. The propagation is synchronized such that output rows receive the correct tag. In the case where $B$ is blinded (not shown here), all output from the systolic array would be blinded since $B$ is preloaded into the array.
    }
    \label{fig:propagation}
\end{figure*}

The most straightforward way to perform \gls{DIFT} is to insert tracking logic inside every \gls{PE}. However, this introduces unnecessarily large overhead. Instead, we implement \gls{DIFT} at row granularity. This fits well with how Gemmini operates, as explained in \Cref{sec:bg:gemm}, and, in weight-stationary mode, enables inputs from different clients to be streamed back-to-back into the systolic array while maintaining isolation between them. We analyze the systolic behavior of the tiles in Gemmini and deduce how tags will propagate from inputs to outputs, which is possible due to the fixed functionality of each tile. We calculate the tags for each output row according to our \gls{DIFT} policy before the inputs enter the systolic array, and then propagate the tags using a minimal circuit in parallel to the data. This allows us to avoid adding logic to each tile and \gls{PE}. The propagation of output tags is visualized in \Cref{fig:propagation}. In this example, the parallel circuit allows us to reduce tag storage registers from four (one within each tile) to three. It also reduces tag propagation logic from 4xNxM instances, where N and M are the dimensions of the \gls{PE} array within each tile, to only three. These reductions scale quadratically with larger systolic arrays. We show how these reductions benefit larger configurations in \Cref{sec:eval}.

Each matrix row carries a single blindedness tag, where a zero tag indicates non-blinded, and every non-zero tag value identifies a separate security domain.
Output rows always take a constant number of cycles to propagate out of the systolic array. We therefore propagate the tag using a parallel queue with a latency that matches that of the output rows. The queue input tag is determined by an OR of the tags corresponding to the three input matrices. This is done after checking that the three inputs do not have mismatching non-zero tags, as explained in \Cref{sec:mixing}.

\subsection{Scratchpads \& context switches}
The internal scratchpads are extended with tags to mark secret data inside Gemmini. This is necessary because the \gls{OS} is in charge of context switches. If the internal scratchpads are not tagged, a malicious \gls{OS} can wait for a process to load secret data into Gemmini, and then perform a context switch to give another process (or itself) access to the data. We extend the scratchpads with one tag per entry, which is equal in size to the rows of the systolic array. As in \Cref{sec:dift}, this corresponds to one tag per row and therefore, all elements in the row share the same tag.

\subsection{Tag mixing}\label{sec:mixing}
We do not allow mixing of data with different non-zero tags. Therefore, when calculating the tag for each output row, we check the tags of the three corresponding inputs (rows of $A$ and $D$, and column of $B$) and only allow propagation when there are no two or more different non-zero tags. Furthermore, when reading data from main memory and writing it to the scratchpads, mismatching tags can occur if the width of the scratchpad rows is larger than 64 bits. We therefore also check for and forbid mismatching tags between the current tag of the scratchpad row and the tag of incoming data.

\subsection{Read-Check-Write}\label{sec:readcheckwrite}
\begin{figure}[ht]
    \centering
    \includegraphics[trim={4.4cm 3cm 3cm 3cm},clip,width=\linewidth]{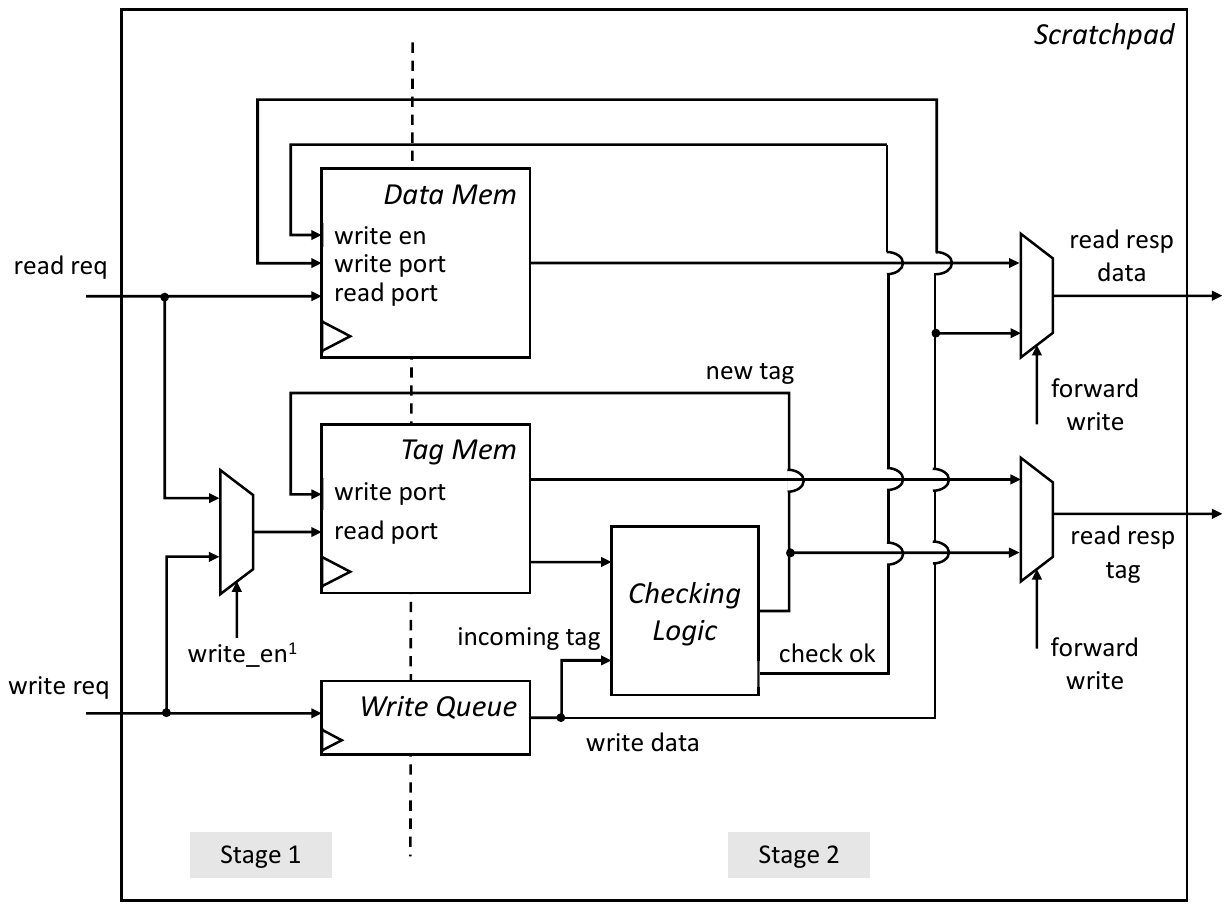}
    \caption{Writes are pipelined to maintain throughput while enabling read-check-write to prevent tag mixing. A 2-stage pipeline is implemented. The dashed line visualizes the separation between the two stages.}
    \label{fig:pipelinedWrites}
\end{figure}

As discussed in \Cref{sec:mixing}, partial writes require first checking that there is no violation between the current and incoming row tags. However, scratchpad memory is implemented using SRAM, which provides a synchronous read/write interface and requires that reads and writes take one cycle. This means that we cannot read the current tag, check for violations, and then request a write all in one cycle. A simple solution to this would be to implement the tag memory (only) using registers to support asynchronous reads. However, this is infeasible due to the large capacity of the scratchpads leading to high area overhead. We therefore pipeline the writes, introducing an additional cycle to read the current tag and delaying the write to the second pipeline stage. This is shown in \Cref{fig:pipelinedWrites}. The scratchpad processes a maximum of one read or write per cycle (not both), with priority given to writes. For a read in Stage 1, we issue corresponding reads to the data and tag memories, which are available for output after one cycle at Stage 2. For a write in Stage 1, we issue a read to the tag memory to check the current tag in memory. In Stage 2, we output read responses or issue writes to memory after checking for violations. In the case that the scratchpad receives a write followed by a read to the same address, this would mean that we issue a read (in Stage 1) and a write (in Stage 2) to the same address \emph{in the same clock cycle} (i.e., a Read-Over-Write). This is not supported by the underlying SRAMs. We therefore manually implement a bypass to forward writes to reads. Data is only forwarded if no violations are detected in the corresponding write.

\subsection{Activation functions}
One important aspect of accelerator architectures is that common activation functions are implemented in hardware using combinational logic. This is in contrast to general-purpose architectures where activation functions are normally computed in software. Computation of activation functions, such as ReLU, in software is often implemented using branching logic, which leads to timing side channels. This is because the control flow becomes dependent on the data values (\Cref{sec:sidechannels}). For example, in the case of ReLU, whether to multiply by a constant factor or set to zero depends on whether the input is positive or negative. By implementing the functions using combinational logic, the function becomes constant-time for all inputs, eliminating all timing side channels. This allows us to simply propagate taint from inputs to outputs without any cycle overheads.

\section{Evaluation}\label{sec:eval}

\subsection{Performance, power \& resource usage}

\subsubsection{Performance}\label{sec:performance}

Performance overheads consist of two parts: cycle overhead, and clock frequency overhead. Cycle overhead is caused by the additional clock cycles required to access the tag bits in main memory and by the added latency for read-check-write (\Cref{sec:readcheckwrite}). Clock frequency overhead is caused by the additional circuitry, which can cause a reduction in maximum clock frequency if it is on the critical path.

\begin{figure}[ht]
    \centering
    \includegraphics[width=0.98\linewidth]{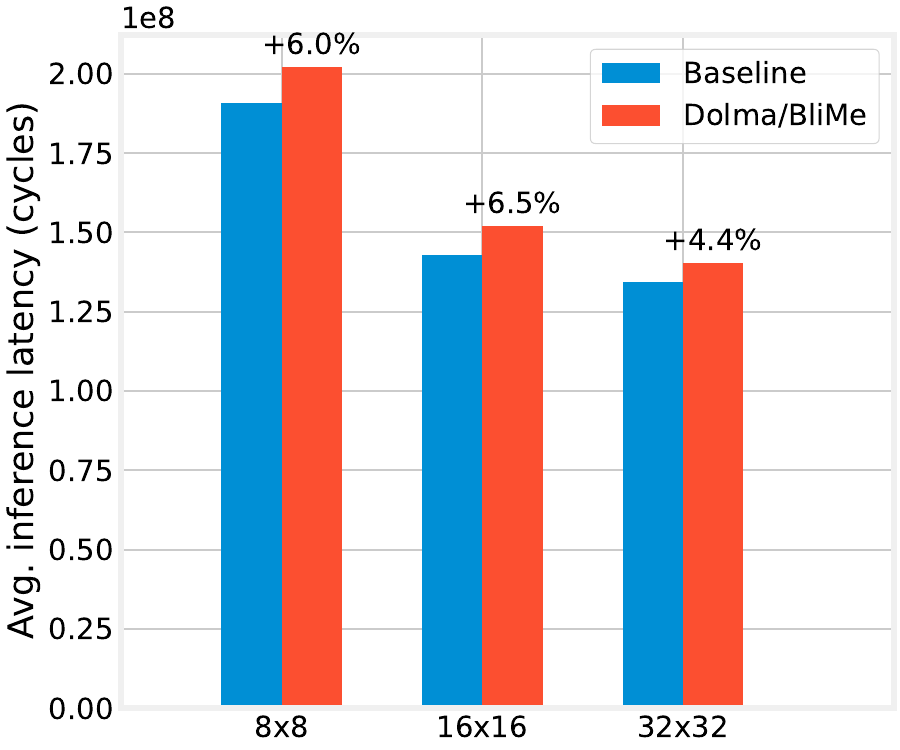}
    \caption{Performance results of running ResNet-50 image classification for \design{} relative to unmodified Gemmini. We obtain an average overhead of 5.6\% over all three configurations.}
    \label{fig:perfeval}
\end{figure}

We measure cycle overhead by running \gls{ML} inference in the form of image classification on ResNet-50 using a random 100-image sample of the ImageNet data set. The experiments were executed on a Xilinx VCU118 FPGA running Linux Buildroot on the Chipyard design, which was configured with a single BOOM core. We compare the performance of \design{} to baseline Gemmini without BliMe. The results are shown in \Cref{fig:perfeval}. We measure an average overhead of 5.6\%. This is similar to overhead reported by BliMe~\cite{blime24} compared to unmodified BOOM, indicating that the majority of the overhead is not inherent to Gemmini, but rather to the memory tagging mechanism used by BliMe. 

All configurations for both \design{} and the baseline were successfully built for the default Chipyard VCU118 frequency of 50MHz. Xilinx Vivado reported baseline \gls{WNS} (in $ns$) as 0.263, 0.367 and 0.261 for the 8x8, 16x16 and 32x32 configurations, respectively. \Gls{WNS} for \design{} was 0.309, 0.300 and 0.166, respectively. This presents no significant changes and indicates that the \gls{DIFT} logic is not on the critical path.

\subsubsection{Power}\label{sec:power}
We obtain Vivado power reports for \design{} and unmodified Gemmini for all configurations. The results are shown in \Cref{tab:power}, with estimated power consumption increasing by 12.2$-$16.5\%.

\begin{table}[ht]
\centering
\begin{tabular}{l|r|r|r}
\multirow[b]{2}{*}{\textbf{Config}} & \multicolumn{3}{c}{\textbf{Power consumption (\si{\watt})}} \\ \cline{2-4}
& \textbf{Baseline} & \textbf{\design{}} & $\Delta\;[\%]$  \\
\hline
8x8   & 4.951    & 5.701 & 15.2 \\
16x16 & 5.426    & 6.085 & 12.2 \\
32x32 & 6.994    & 8.147 & 16.5
\end{tabular}
\vspace{0.5em}
\caption{Effect of modifications on FPGA power consumption.}
\label{tab:power}
\end{table}

\subsubsection{Resource usage}
\begin{figure}[ht]
    \centering
    \includegraphics[width=0.98\linewidth]{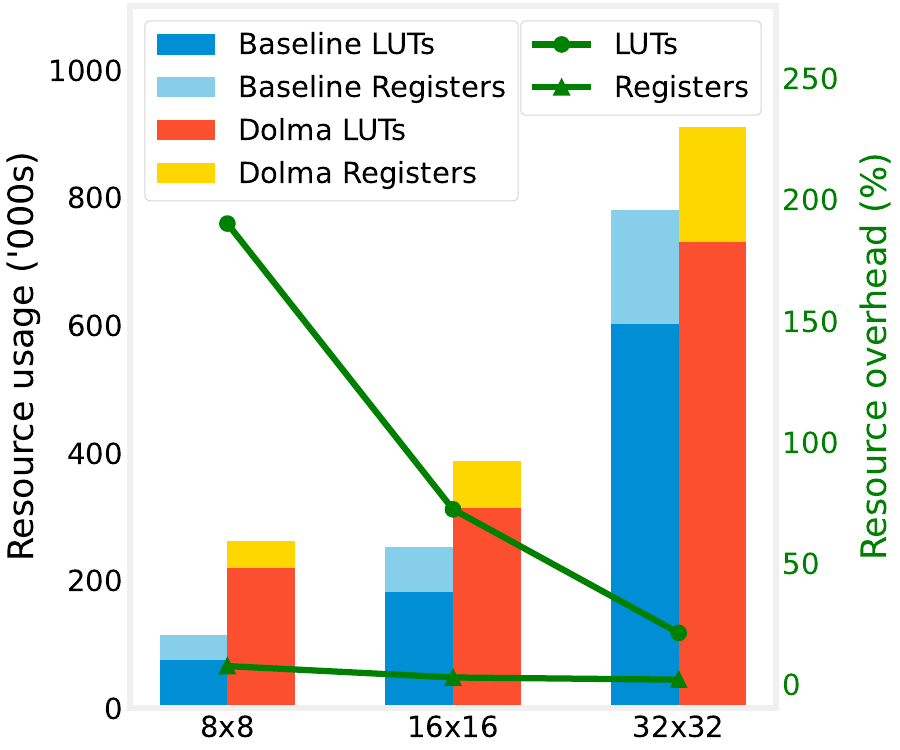}
    \caption{Resource usage results for \design{} relative to unmodified Gemmini.}
    \label{fig:resourceeval}
\end{figure}

We obtain FPGA resource usage reports from Vivado for \design{} and unmodified Gemmini for different mesh sizes. The results are shown in \Cref{fig:resourceeval}. The results here highlight the benefits of the optimization described in \Cref{sec:dift}. We see a significant decrease in relative resource usage overheads as the mesh dimensions increase. Note that the additional resource usage reported here includes the tag storage added to the scratchpads, whose size is constant across the three configurations.

\subsection{Security}\label{sec:security}

BliMe includes a formal model and a machine-checked security proof in F*~\cite{fstar}. It proves that safe computation, as defined by BliMe, does not leak any information on data marked as blinded.  However, \design{} differs from normal instructions in that it directly modifies memory, rather than being limited to data stored in registers, and we therefore extend BliMe's formal model to include this type of operation.  In particular we prove that adding `safe' accelerator operations to the system also does not leak any information.

First, we define \gls{DMA} accelerator operations as functions that map a list of input words (read from memory) to a list of outputs (written to memory). We then define `safe' accelerator operations as those that do not allow for information to flow from their blinded inputs to their observable outputs; more specifically, if two inputs to the accelerator differ only in their blinded values, then their corresponding outputs can differ only in their blinded values.

We modify the execution model to allow execution units to trigger such an accelerator, and prove that any accelerator meeting this safety requirement does not compromise the safety of the system. We then modify BliMe's concrete instruction set architecture model to incorporate a model \design{} in the form of a matrix multiply-and-add accelerator, which if given any blinded input produces a fully blinded output.  This accelerator can be proven safe, thereby showing that incorporating our model \design{} into BliMe's model \gls{ISA} does not undermine its security guarantees.

\section{Discussion \& Future Work}

\subsection{Data-dependent processing}
Gemmini does not introduce any data-dependent control flow. As a result, the only potential for policy violation arises from tag mixing (\Cref{sec:mixing}). However, state-of-the-art accelerators often provide data-dependent optimizations, e.g., skipping multiplications when zeros are detected is useful when processing sparse matrices~\cite{Sze2017EfficientPO}. These can result in data-dependent computation latencies leading to data leakage through timing side channels. Furthermore, some accelerator designs enable on-chip control flow~\cite{tpu}. This allows different control paths to be taken depending on the data, which can again leak data through side channels as described in \Cref{sec:sidechannels}.
Implementing \design{} on such accelerators must consider \emph{all} data-dependent timing side channels and enforce data-obliviousness when processing blinded data. To avoid unnecessarily enforcing worst-case data-oblivious performance on all data, additional logic will likely be needed to disable optimizations on blinded data \emph{only}.

\subsection{GPUs}
\Glspl{GPU} are another type of accelerators that are commonly used for \gls{ML} workloads. Originally intended for graphics processing, \glspl{GPU} contain a large number of small cores that execute small sequences of code called ``kernels'' in parallel. After the introduction of general-purpose \glspl{GPU}, which enable arbitrary code execution, \gls{GPU} architectures have recently shifted towards incorporating dedicated hardware components for accelerating \gls{ML} workloads~\cite{nvidiaTensorCores}.

In contrast to accelerators such as Gemmini, \glspl{GPU} inherently support on-chip control flow for kernel execution, making them vulnerable to control-flow-based timing side channels. As discussed in the previous section, this requires additional care to enforce data-obliviousness on branching instructions. Another source of timing side channels is memory accesses that use secret-dependent addresses. Memory accesses to off-package main memory are subject to the same data-oblivious policy, i.e., ones that use secret-dependent addresses are forbidden. However, \glspl{GPU} use high-capacity on-package memories and provide explicit control over data movement in and out of these memories using a separate local address space (unlike \gls{CPU} caches). This means that secret-dependent accesses to these on-package memories (using blinded local addresses) can be executed without causing timing differences visible to an adversary, providing opportunities for optimizing data-oblivious algorithms. On the other hand, \glspl{GPU} are more general-purpose than fixed accelerators such as Gemmini and might not be able to benefit from row-wise \gls{DIFT}. We leave such research as future work.

\section{Related work}

There is an extensive body of research on \gls{DIFT}. Hu \emph{et al}.~\cite{huHardwareInformationFlow2021} have conducted a survey encompassing numerous hardware-based taint-tracking approaches. These techniques pursue diverse objectives and involve trade-offs between security and performance, operating at different levels of abstraction. Tiwari \emph{et al}.~\cite{GLIFT} suggest GLIFT, which applies \gls{DIFT} at the gate level, employing it to design a processor capable of monitoring all information flows. However, this suffers from significant overheads. \design{}, on the other hand, applies \gls{DIFT} at a larger granule and forbids explicit and implicit flows from secret user data to control data (e.g., program counter) and instructions.

Several \gls{DIFT} processor designs exist. The type of metadata tracked by the processor and the policies enforced on them can be fixed or programmable. HardBound~\cite{hardbound2008} uses a fixed policy where the semantic meaning of the tags and the rules governing them are hardwired. Each tag represents whether the corresponding word is a pointer, and the hardware uses the tag to determine whether it should apply its bound-checking policy. On the other side of the spectrum, PUMP~\cite{pump2014} and Raksha~\cite{raksha2007} provide support for user-defined policies, which allows assigning arbitrary meaning to tags. However, both perform the policy enforcement step late in the processor pipeline. This makes them unsuitable for preventing side-channel leakage as memory side effects (e.g., fetching sensitive addresses into the cache) occur before the policy is enforced. HyperFlow~\cite{hyperflow2018} takes programmable policies a step further by providing support for timing side channels. The authors introduce a new secure \gls{HDL} called ChiselFlow which they use to implement the processor. However, this requires significant source code and design changes in order to pass the ChiselFlow secure compilation step. Furthermore, HyperFlow uses a trusted \gls{OS} to assign security labels to processes and control information-flow policies between processes.

\section{Conclusion}

We introduced \design{}, a minimal extension to matrix multiplication hardware accelerators that enables efficient \gls{DIFT} and guarantees confidentiality of secret client data in outsourced computation. Efficient \gls{DIFT}, in general, can be achieved by identifying fixed-function components in a system and removing unnecessary tracking logic within those components. \design{} implements this by identifying the fixed functionality of systolic arrays and applying row-wise tracking logic. Our results show this to be be an attractive solution for enabling efficient \gls{DIFT} in fixed-function computing units, which can be extended to other units such as cryptographic accelerators and \glspl{GPU}.

\bibliographystyle{IEEEtran}
\bibliography{IEEEabrv,refs}

\end{document}